\documentstyle[12pt]{article}

\newcommand{\be}{\begin{equation}}
\newcommand{\ee}{\end{equation}}
\newcommand{\bea}{\begin{eqnarray}}
\newcommand{\eea}{\end{eqnarray}}
\newcommand{\beaa}{\begin{eqnarray*}}
\newcommand{\eeaa}{\end{eqnarray*}}

\newcommand{\th}{\theta}
\newcommand{\del}{\partial}
\newcommand{\g}{{\cal G}}
\newcommand{\J}{{\cal J}}
\newcommand{\K}{{\cal K}}
\newcommand{\p}{{\cal P}}

\newcommand{\Lg}{{\bf g}}
\advance\textheight by \topskip
\makeatletter

\@addtoreset{equation}{section}
\def\section{\@startsection {section}{1}{\z@}{-3.5ex plus -1ex minus
 -.2ex}{2.3ex plus .2ex}{\large\bf\centering}}
\def\subsection{\@startsection{subsection}{2}{\z@}{-3.25ex plus%
 -1ex minus -.2ex}{1.5ex plus .2ex}{\bf}}
\def\subsubsection{\@startsection{subsubsection}{3}{\z@}{-3.25ex plus%
 -1ex minus -.2ex}{1.5ex plus .2ex}{\sl}}
\makeatother

\begin{document}

\baselineskip 16pt 
\parindent 12pt
\parskip 8pt 

{
\parskip 0pt
\newpage
\begin{titlepage}
\begin{flushright}
DAMTP-97-120\\
hep-th/9711140v3\\
Revised January 1999\\[3mm]
\end{flushright}
\vspace{.4cm}
\begin{center}
{\Large {\bf
Conserved Charges and Supersymmetry \\in Principal Chiral
Models\footnote{Revised and up-dated version based on talks 
by NJM at the 3rd Trieste Conference on Statistical Field Theory, 
June 1998, and by JME at the 2nd Annual TMR 
Network Conference: Integrability, Non-perturbative Effects, and Symmetry in
Quantum Field Theory, Durham, September 1998.}
}}\\
\vspace{1cm}
{\large J.M. Evans${}^a$\footnote{E-mail: J.M.Evans@damtp.cam.ac.uk,
M.U.Hassan@damtp.cam.ac.uk, N.MacKay@sheffield.ac.uk, 
A.Mountain@ic.ac.uk} 
M. Hassan${}^{a 2}$ N.J. MacKay${}^{b 2}$ A.J. Mountain${}^{c 2}$
}
\\
\vspace{3mm}
{\em ${}^a$ DAMTP, University of Cambridge, Silver Street, Cambridge
CB3 9EW, U.K.}\\
{\em ${}^b$ Department of Applied Mathematics, 
University of Sheffield, Sheffield S3 7RH, U.K.}\\
{\em ${}^c$ Blackett Laboratory, Imperial College, Prince Consort Road,
London SW7 2BZ, U.K.}\\

\vspace{1.5cm}
{\bf{ABSTRACT}}
\end{center}
\begin{quote}{\small
We report on investigations of local (and non-local) charges in bosonic and
supersymmetric principal chiral models in 1+1 dimensions. 
In the bosonic PCM there is a classically conserved local charge 
for each symmetric invariant tensor of the underlying group.
These all commute with the non-local Yangian charges. 
The algebra of the local charges amongst themselves is rather
more subtle. We give a universal formula for infinite sets of mutually
commuting local charges with spins equal to the exponents of the 
underlying classical algebra modulo its Coxeter number.
Many of these results extend to the supersymmetric PCM, but 
with local conserved charges associated with antisymmetric invariants 
in the Lie algebra.
We comment briefly on the quantum conservation of local charges
in both the bosonic and super PCMs.}

\end{quote}
\vfill
\end{titlepage}
}

\section{Introduction}

Integrable QFTs in two spacetime dimensions display many of
the most important phenomena of higher-dimensional QFTs in a 
tractable setting. Non-linear sigma-models exhibit features such 
as asymptotic freedom, dynamical mass generation,
and confinement; while within the class of Toda theories we find the
famous sine-Gordon model---providing (together with the massive 
Thirring model) the prototype example of an exact equivalence 
in which perturbative and solitonic degrees of freedom are exchanged.
Our understanding of such striking phenomena is strengthened 
by our ability to find exact S-matrices for many of these 
two dimensional theories.
As well as a bootstrap principle and standard axioms like
analyticity and unitarity, these S-matrices are highly constrained 
by the existence of `exotic' conserved quantities,
so that a multi-particle process must factorize into two-particle
scatterings, and these in turn must obey the Yang-Baxter equation.
It is the requirements of factorization and elasticity, 
above all else, which allows the S-matrix to be determined.
The nature and properties of the `exotic' charges which lie behind
this can vary greatly from model to model, however. 

We report here on results \cite{EHMM} for a particular class of non-linear
sigma-models (and their supersymmetric extensions) with target manifold
a compact Lie group: these are the
principal chiral models or PCMs.
In addition, much of what we do is naturally viewed against 
progress in understanding affine Toda
field theories or ATFTs (though we shall not discuss 
them in any great detail beyond this introduction;
see {\it e.g.}~\cite{corri94} for a review). 
ATFTs are also defined by Lie algebra data, 
this time a set of simple roots which specify the exponential
interactions amongst a set of scalar fields.
The two classes of models (PCMs and ATFTs) exemplify the 
very wide range of physical and mathematical behaviour encountered 
in integrable QFTs.

ATFTs can be treated 
in the classic perturbative manner, using Feynman diagrams to
calculate order-by-order from the classical lagrangian after 
identifying the interactions among classical mass eigenstates. 
Such calculations confirm and complement the exact formulas for the 
S-matrices \cite{brade90,corri94}. 
The PCMs, in contrast, have quantum 
couplings which become very large at low energies, so that many  
quantum properties are impossible to extract analytically from the 
classical lagrangian. Under these circumstances the S-matrices 
\cite{ogie86,evans96} must 
be checked by other means (see {\it e.g.}~\cite{TBA} and reference 
therein).

Another important difference concerns the nature of the conserved
quantities responsible for factorization.
ATFTs have infinitely many {\em local\/} conserved charges 
({\it i.e.}~integrals of local densities) which commute with one another.
These are `exotic' only in as much as they have higher-spins
(and would therefore be forbidden in 
four dimensions by the Coleman-Mandula Theorem)
with values running over the exponents of the underlying Lie algebra
modulo its Coxeter number (\cite{corri94} and references therein). 
The lowest exponent is always one, and the lowest-spin charge is thus
energy-momentum. The conservation of these charges in particle fusings 
({\it i.e.}~three-point couplings 
in perturbation theory and certain S-matrix poles in exact scattering)
can be characterised by an elegant geometrical construction known as
Dorey's rule \cite{dorey91,corri94}.
Each fusing is described by a triangle in the higher-dimensional space
of roots, with the values of the individual charges being obtained by
projecting this down onto a canonical set of planes through root space.

The PCMs have conserved charges which are much more `exotic' 
from a conventional QFT standpoint: {\em non-local\/} quantities 
with an associated quantum group structure known as a Yangian 
\cite{qyang,cyang}. 
(ATFTs exhibit a related quantum
group structure in the more complicated case when their coupling is
imaginary, but we shall not dwell on this here---see \cite{ganden96}
for a review.)
These charges have non-integer or even indefinite spin, and, because 
they are the integrals of non-local densities, a non-trivial addition
rule (coproduct) on asymptotic states. Their non-vanishing commutation
relations may be regarded as a spacetime extension of internal
Lie algebra symmetries. 

Despite the profound differences between the local and non-local
charges, they turn out to have a surprising commonality of
features and consequences. One of these is the basic property of
factorization of the S-matrix itself, which the two kinds of charges
enforce in quite different ways.
Another point of specific relevance to the work summarized here 
is the important result of 
Chari and Pressley \cite{chari95}: that
the Yangian quantum group fusing rule
for particle multiplets in the PCM,
and Dorey's rule describing fusings for particles in ATFTs,
are one and the same. 

This naturally suggests some deeper underlying connections 
in integrable QFTs, and in particular it begs the
questions of when local and non-local charges co-exist, and what 
their relationship might then be.
In fact the existence of local conserved charges 
in PCMs has been known for some time, but they have received 
comparatively little attention. Our work \cite{EHMM}
has led to a general construction of infinitely many local
commuting charges in each PCM based on a classical algebra,
with spins equal to the exponents modulo the Coxeter number, exactly
as for ATFTs. We have also begun to investigate 
the analagous questions for the supersymmetric PCMs.
These are known to have some novel features and are generally much 
less well-understood than their bosonic counterparts.
\pagebreak

\section{The classical bosonic PCM}

\subsection{The model in outline}
The principal chiral model is defined by a field $g(x)$
taking values in a compact Lie group ${\cal G}$ and governed by the 
lagrangian\footnote{Spacetime conventions: orthonormal 
and light-cone coordinates are related by 
$x^\pm = {1\over 2} (t \pm x)$ and $\del_\pm = \del_t \pm \del_x$.
Lie algebra conventions:
We take $\Lg$ in its defining representation
with anti-hermitian generators $t^a$ obeying
$[ t^a , t^b ] = f^{abc} t^c$ and ${\rm Tr} (t^a t^b) = -\delta^{ab}$.
For $X \in {\Lg}$ we write 
$X = t^{a}\, X^a$ and $X^a = - {\rm Tr}(t^a X)$.}
\be\label{pcmlagr}
{\cal L} = {1 \over 2 } {\rm Tr}\left( \partial_+ g^{-1}
\partial_- g\right) \, .
\ee
There is a global symmetry 
${\cal G}_L \times {\cal G}_R$ under which 
$g \mapsto U^{\phantom{1}}_{\rm L} \,g\, U^{-1}_{\rm R}$.
The current 
\be\label{lrcurr} 
j_\pm = - g^{-1}\partial_{\pm} g
\ee
takes values in the Lie algebra $\Lg$ and 
corresponds to ${\cal G}_R$ transformations, while 
$- g j_\pm g^{-1}$ corresponds to 
${\cal G}_L$ transformations.
The equations of motion for the PCM are 
\be\label{lce}
\partial_- j_+ = - \partial_+ j_- = - {1\over 2} [j_+,j_-] \, .
\ee
The energy-momentum tensor has components 
\be\label{conf} 
T_{\pm \pm} = -{1\over 2 } {\rm Tr} ( j_{\pm} j_{\pm}) , 
\qquad T_{+-} = T_{-+} = 0 , \qquad \mbox{with} \qquad
\partial_- T_{++} = \partial_+ T_{--} = 0 \, ,
\ee
reflecting the classical conformal symmetry of the theory.

Every PCM has an important discrete symmetry
$\pi : g \mapsto g^{-1}$ which exchanges $\g_L$ and $\g_R$.
Other discrete symmetries arise as outer automorphisms of $\g$ acting 
on the field $g$.
Thus we have a new symmetry $\gamma : g \mapsto g^*$ 
when the defining representation of $\Lg$ is complex,
while for $\Lg = so (2 \ell)$ we also have 
$\sigma: g \mapsto MgM^{-1}$ (where ${\rm det}M{=}-1$) 
which exchanges the inequivalent spinor representations.
Because higher derivatives of $j_\pm$ do not transform in a
simple fashion under these discrete symmetries, it is
convenient to introduce 
$$
j_{++} \equiv \partial_+ j_+ \,, \quad
j_{+++}\equiv \partial_+ j_{++}\, -\, {1\over 2} [j_+,j_{++}] \,, \quad
j_{++++}\equiv \partial_+ j_{+++}\, -\, {1\over 2} [j_+,j_{+++}] \,,
\quad \ldots
$$
(and similarly for the the minus components).
It is then easy to show  
\be
\pi \;: \; 
j_{++\ldots+} \; \mapsto \; - g \, j_{++\ldots +}\, g^{-1} \; ,
\qquad 
\gamma \;: \; 
j_{++\ldots+} \; \mapsto \; (j_{++\ldots +})^* = - (j_{++\ldots +})^T 
\ee
which will be useful later.

The canonical structure of the classical PCM is defined by
the Poisson brackets of the currents: 
\begin{eqnarray}
\left\{ j_\pm^a (x), j_\pm^b (y) \right\} 
& = & f^{abc} ( \, {\textstyle { 3 \over 2}} j_\pm^c (x) - 
{\textstyle {1 \over 2}} j_\mp^c (x) \, ) \,
\delta(x{-}y) \, \pm \, 2 \delta^{ab} \delta'(x{-}y) \nonumber \\
\left\{ j_+^a (x), j_-^b (y) \right\} 
& = & {\textstyle {1\over 2}} \, f^{abc} (\, j_+^c (x) +j_-^c (x) \, )\, 
\delta(x{-}y) 
\label{PBs} 
\end{eqnarray}
at equal time.
These imply that the energy momentum tensor satisfies the 
classical, centre-less Virasoro algebra 
\be\label{vir}
\{ T_{++} (x) , T_{++} (y) \} =  2 \delta'(x{-}y) 
( \, T_{++}(x) + T_{++} (y) \, )
\ee
(and similarly for the minus components).

\subsection{Local conserved charges}
There are several categories of higher-spin, conserved, local charges
in the PCM which have distinct characteristics. 

\noindent $\bullet$ 
The conservation of the energy-momentum tensor (\ref{conf}) 
immediately implies 
\be\label{confpower}
\partial_- (T^n_{++}) = \partial_+ (T_{--}^n) = 0 \, . 
\ee
Such conservation laws clearly hold in {\em any\/} classically 
conformally-invariant theory.

\noindent $\bullet$
In addition, we have
\be\label{curr}
\partial_- {\rm Tr} (j_{+}^m) = 
\partial_+ {\rm Tr} (j_{-}^m) = 0 \; .
\ee
These are a consequence of (\ref{lce}); they  
depend on the detailed form of the PCM equations of motion,
rather than on conformal invariance alone.

\noindent $\bullet$
The two previous categories may now be generalized as follows.
Let $d_{a_1 a_2 \ldots a_m}$ be any totally symmetric
invariant tensor, so that 
$d_{c(a_1a_2 \ldots a_{m-1}}f_{a_m)bc} = 0$. 
For each such tensor (or Casimir) there are conservation equations
\be 
\label{gencons}
\del_-( \, d_{a_1 a_2 \ldots a_m} j_+^{a_1} j_+^{a_2} \ldots
j_+^{a_m} \, ) = 
\del_+( \, d_{a_1 a_2 \ldots a_m} j_-^{a_1} j_-^{a_2} \ldots
j_-^{a_m} \, ) = 0 \ .
\ee
The currents in (\ref{confpower}) correspond to  
even-rank invariant tensors constructed from Kronecker deltas:
\be \label{sdelta}
d_{a_1a_2 \ldots a_{2n-1} a_{2n}} = 
\delta_{(a_1 a_2} \delta_{a_3 a_4} \! \ldots \delta_{a_{2n-1} a_{2n} )} 
\ee
while those in (\ref{curr}) correspond to 
\be \label{strace}
d_{a_1a_2 \ldots a_m} = 
{\rm STr}(t^{a_1} t^{a_2} \! \ldots t^{a_m}) 
\ee
with `STr' denoting the trace of a completely symmetrized product of 
matrices.

\noindent
$\bullet$ Finally, the most general possibility is to take an
arbitrary {\em differential\/} polynomial in the 
currents we have already discussed,
{\it e.g.\/} $\del_- \left ( {\rm Tr} (j_+^p ) \del_+^r {\rm Tr}
(j_+^q) \right ) = 0$ follows immediately from (\ref{curr}).
We shall not be directly concerned with 
such currents here. 

These observations lead us to a more detailed consideration of
invariant tensors. There are infinitely many invariant tensors for each 
algebra $\Lg$ but only $rank(\Lg)$ independent or {\it primitive\/} 
$d$-tensors and Casimirs (see {\it e.g.}~\cite{azca97}), whose degrees
equal the exponents of ${\Lg}$ plus one. All other invariant tensors 
can be expressed as polynomials in these and the structure constants 
$f_{abc}$. The primitive $d$-tensors for the classical algebras can
all be chosen to be symmetrized traces, as in (\ref{strace}),
with one exception. This exception is the Pfaffian invariant for 
$so(2\ell)$, which has rank $\ell$, and can be written
\be\label{pfaff}
d_{a_1...a_\ell} = \epsilon_{i_1j_1 \ldots i_\ell j_\ell} 
(t^{a_1})_{i_1 j_1} \ldots
(t^{a_\ell})_{i_\ell j_\ell} \, .
\ee

In the next section we will discuss the algebra of conserved charges 
arising from (\ref{gencons}) for various choices of the invariant 
tensors $d$. We denote these charges 
\be \label{locch}
q_{\pm s} \,=\, d_{a_1a_2 \ldots a_m}\int_{-\infty}^\infty 
j_\pm^{a_1}(x) j_\pm^{a_2} (x) \ldots j_\pm^{a_m}(x)
\,dx 
\ee
labelled by their spin, $s= m{-}1$. It will be sufficient for many 
purposes to consider $q_s$ with $s>0$, and it is useful to introduce 
the notation 
\be\label{simple}
\J_m = {\rm Tr} (j_+^m) \, , \qquad 
\p_{\ell} = \epsilon_{i_1 j_1 \ldots i_\ell j_\ell} (j_{+})_{i_1 j_1} \ldots
(j_{+})_{i_\ell j_\ell}
\ee
for the currents corresponding to the tensors (\ref{strace}) and
(\ref{pfaff}). Once again, sub-scripts denote the spin.

\subsection{Non-local conserved charges}
In addition to the local charges,
there exist infinitely many conserved non-local charges in
the bosonic PCM, generated by the obvious local charge 
$$
Q^{(0)a}  =  \int^{\infty}_{-\infty} j_{0}^{a}(x) {dx} 
$$
and the rather less obvious first non-local charge
$$
Q^{(1)a}  =  \int^{\infty}_{-\infty} j_{1}^{a} (x) {dx} - 
{1\over 2}f^{abc}\int^{\infty}_{-\infty} j_{0}^{b}(x) 
\int_{-\infty}^{x} j_{0}^{c}(y) \,{dy} \,{dx} \, .
$$
Under the Poisson bracket, these form a Yangian $Y({\Lg})$ 
\cite{qyang,cyang}. 
In fact there are two infinite sequences of such charges
constructed from both $j_\mu$ and $-gj_\mu g^{-1}$,
and so the model has 
a charge algebra $Y_L({\Lg})\times Y_R({\Lg})$. (It can be
checked that $Y_L$ and $Y_R$ commute.) These charges
can be extracted from the monodromy matrix
by a power series expansion in the spectral parameter, or via an
iterative construction of their currents \cite{cyang} .

The non-local character of the Yangian charges means that 
they will not be additive on products of states in the quantum theory. 
Instead their action is given by the co-product rules 
$\Delta ( Q^{(0)a} ) = Q^{(0)a} \otimes 1 + 1 \otimes Q^{(0)a} $,
which is essentially trivial,
and $\Delta ( Q^{(1)a} ) = 
Q^{(1)a} \otimes 1 + 1 \otimes Q^{(1)a} + {1\over2} 
f^{abc} Q^{(0)b} \otimes Q^{(0)c}$,
which is non-trivial.
These equations may also be interpreted classically as giving the 
values of the charges on widely-separated, localized configurations 
\cite{qyang,cyang}.

It is natural to ask how the non-local charges behave with respect
to the other conserved quantities in the PCM.
It can be shown \cite{EHMM} that the non-local charges commute
with the general local charge of type (\ref{locch}) in the classical theory:
$\{ q_s , Q^{(0)a} \} = \{ q_s , Q^{(1)a} \} = 0$.
The non-local charges are also classically Lorentz scalars,
as can be checked by applying the boost operator $M$ and calculating
$\{ M, Q^{(0)a} \} =  \{ M, Q^{(1)a} \} =0 $.
A subtle effect in the quantum theory is that the commutator of the 
non-local charges with the boost operator receives a correction 
at $O(\hbar^2)$, which is essential to the non-trivial structure of
the S-matrices for the PCMs. 
Although the calculations have not been carried out, it seems unlikely
to us that the commutators with the charges $q_s$ would receive
similar modifications at the quantum level,
since it is hard to see how this could be compatible with the 
Yangian-invariant S-matrices. Nevertheless, this is something 
which should be checked.

\section{Commuting local charges in the bosonic PCM}

\subsection{Introductory comments and isolated examples}
We have identified several sets of local
conserved charges in any PCM, and an obvious question is whether 
these might be, or might contain, sets which mutually commute.
At this stage one may be rather discouraged
by the form of the current Poisson brackets (\ref{PBs}).
Both the lack of covariance and the presence of the $\delta'$ terms
foreshadow potential complications with the charge algebra.
Indeed, this was sufficient to prompt Faddeev and Reshetikin
\cite{fadd86} to 
introduce and study an alternative classical limit of the quantum PCM
that was more amenable to standard techniques like the classical
inverse-scattering approach.
We will persevere with (\ref{PBs}), however, 
despite these ominous signs. 
We will see that it is 
possible not only to find sets of charges which mutually commute, but also to 
give a general definition: a universal formula valid for all the 
classical algebras.

The classical Poisson bracket algebra of local charges 
$q_{\pm s}$ $(s >0)$ of the general type (\ref{locch}) 
can be calculated from (\ref{PBs}). The terms involving
$\delta(x{-}y)$ ({\em i.e.\ }the {\it ultra-local} terms) 
always vanish by invariance of the $d$-tensors, leaving only 
contributions from the 
$\delta'(x{-}y)$ terms ({\em i.e\ }the {\it non-ultra-local} terms).
It is clear from (\ref{PBs}) that these are absent too 
if we consider charges of opposite chiralities, implying
\be\label{lpb1} 
\{ q_s, q_{-r} \} = 0 \, , \qquad r,s > 0.
\ee 
For charges of the same chirality, however, the result is 
generally non-zero:
\begin{equation}\label{nasty}
\{q_s , q_r \}  =  
(const) \int_{-\infty}^\infty dx\, d_{ca_1 \ldots a_s} 
{d}_{c b_1 \ldots b_{r} } 
j^{a_1}_+ \ldots j^{a_s}_+ 
\del_x ( j^{b_1}_+ \ldots j_+^{b_{r}} ) \,.
\end{equation}
Note that this expression is anti-symmetric in $s$ and $r$, 
by integration by parts. 

Our aim now is to find invariant tensors and conserved currents 
for which this expression vanishes, so that the charges 
commute. 
There are three circumstances in which this happens
in a relatively simple way.

\noindent
$\bullet$ 
For $r=1$ and ${d}_{bc} = \delta_{bc}$, 
the integrand in (\ref{nasty}) is clearly a total derivative
and the Poisson bracket vanishes. This simply means that {\it all\/} the 
local charges (\ref{locch}) commute with energy and momentum:
they are invariant under translations in space and time. 

\noindent
$\bullet$ 
If {\it both\/} currents are powers of the energy-momentum tensor 
(\ref{confpower}) then the integrand in (\ref{nasty}) can 
be written as a total derivative. 
This is actually a general feature of 
any classically conformally-invariant theory, whose 
energy-momentum tensor obeys the Virasoro algebra
(\ref{vir}). It is a simple consequence of this that the charges
$\int \! T_{++}^n \, dx$ commute with one another.

\noindent
$\bullet$ The currents $\J_m$ defined in (\ref{simple}) give rise to 
commuting charges $\int \J_m \, dx$ for $\Lg = so(\ell)$ or $sp(\ell)$.
For these currents we note that (\ref{nasty}) can be written 
\be\label{nicer}
\{q_s , {q}_r \}  = 
(const) \int_{-\infty}^\infty dx\,{\rm Tr}(t^c j_+^s) 
\, \partial_x 
{\rm Tr}(t^c j_+^{r}) \,.
\ee
which can be simplified using the completeness condition 
$ X = - t^a \, {\rm Tr} (t^a X) $ valid for any $X$ in $\Lg$.
For the orthogonal and symplectic algebras $s$ and $r$ are always odd
(otherwise the currents vanish) and this implies that $(j_+)^r$ 
and $(j_+)^s$ also belong to $\Lg$.\footnote{
If $X$ is in $so(\ell)$ it is a real anti-symmetric matrix, 
and so if $m$ is odd, $X^m$ will also be real and anti-symmetric, 
and hence also in $so(\ell)$. Similar arguments apply to $\Lg = sp(\ell)$.}
The completeness condition then implies that the integrand is 
proportional to $\del_x {\rm Tr} (j_+^{r+s})$ and hence 
the charges commute.

The charges $\int \J_m \, dx$ have more
complicated brackets for $\Lg = su(\ell)$.
In this case the completeness condition holds only for traceless
matrices and this property is of course spoiled by taking powers. 
The result is that (\ref{nasty}) becomes 
\be\label{n-su}
\{ q_s , q_r \} = (const)
\int_{-\infty}^\infty dx\,{\rm Tr}(j_+^s) \,\partial_x 
{\rm Tr}(j_+^{r}) 
\ee
which is non-zero in general.
Notice that the bracket nevertheless produces 
a conserved quantity which we recognize, namely a differential
polynomial in the currents $\J_m$.

\subsection{A general construction}
We now investigate the possibility of more 
general sets of commuting charges in the PCM based on any classical
group $\g$. One way to
begin is to carry out some trial calculations 
for low-lying values of the spin.
We can search systematically 
for polynomials $\K_{s+1}(\J_m)$ which are homogeneous in the spin 
and which will give commuting charges $q_s = \int \K_{s+1} \, dx$.

For $\Lg = su(\ell)$ we find, after 
some laborious calculations, the following 
expressions: 
\begin{eqnarray}
\K_2 & = & \J_2 
\nonumber\\
\K_3 & = & \J_3 
\nonumber\\
\K_4 & = & \J_4 - {3 \over 2\ell} \, \J_2^2 
\nonumber\\
\K_5 & = & \J_5 - {10\over 3\ell} \, \J_3\J_2 
\nonumber\\
\K_6 & = & \J_6 - {5\over 3\ell} \, \J_3^2 
- {15\over 4\ell} \J_4\J_2 + {25\over 8\ell^2} \, \J_2^3 
\label{suN}
\end{eqnarray}
These are the unique combinations (up to overall constants) for which 
the corresponding charges commute.
For $\Lg = so(\ell)$ or $sp(\ell)$ similar calculations reveal 
a family of currents with a single free parameter $\alpha$.
The first few examples are: 
\begin{eqnarray}
\K_2 & = & \J_2
\nonumber\\
\K_4 & = & \J_4 - {1\over 2} (3\alpha) \, \J_2^2 
\nonumber\\
\K_6 & = & \J_6 - {3\over 4}(5\alpha) \, \J_4 \, \J_2
+{1\over 8} (5\alpha)^2 \J_2^3 
\nonumber\\
\K_8 & = & \J_8 -{2\over 3}(7\alpha) \, \J_6 \, \J_2
-{1\over 4}(7\alpha) \, \J_4^2 +
{1\over 4}(7\alpha)^2 \, \J_4 \, \J_2^2
-{1\over 48} (7\alpha)^3 \, \J_2^4 
\label{soN}
\end{eqnarray}
Notice that 
this one-parameter family 
interpolates the two simplest families we found previously for the
orthogonal and symplectic algebras.
When $\alpha \rightarrow 0$ we have $\K_{2m} \rightarrow \J_{2m}$ 
and in the limit 
$\alpha \rightarrow \infty$ we have (with a suitable rescaling)
$\K_{2n} \rightarrow (\J_2)^n$. 

The examples we have just given are sufficient to suggest general 
definitions of infinite sets of currents which we can then prove 
yield commuting charges.
For each of the classical algebras $su(\ell)$, $so(\ell)$, $sp(\ell)$,
we introduce the generating functions $A(x,\lambda)$ and
$F(x,\lambda)$ defined by 
\be\label{Agen}
A (x , \lambda) = \exp F(x, \lambda) 
= \det ( 1 - \lambda j_+ (x) ) 
\ee
so that 
\be\label{Fgen}
F(x , \lambda) = {\rm Tr} \log (1 - \lambda j_+(x) ) 
= - \sum_{r=2}^\infty {\lambda^r \over  r} \J_r (x) \; . 
\ee
Observe that $A(x , \lambda)$ is a polynomial of degree at most $\ell$,
with the coefficient of the term in $\lambda^\ell$ being 
$(-1)^\ell {\rm det} (j_+)$; on 
substituting the series expansion for $F$ into (\ref{Agen}),
we obtain non-trivial identities satisfied by the $\J_m$
as the coefficients of $\lambda^r$ must vanish for 
$r > \ell$ (for details see {\it e.g.}~\cite{azca97}). 

We now define 
\be\label{def}
{\cal K}_{s+1} = A(x,\lambda)^{\alpha s } \; \; {\Big |}_{ \lambda^{s+1} } 
= \exp {\alpha s } F(x, \lambda) \; \; {\Big |}_{ \lambda^{s+1} }  
\ee
In extracting the coefficients indicated, 
we expand the generating function in ascending powers of $\lambda$.
It can be shown from this definition that the charges 
defined commute when  
$\alpha = 1/h  = 1/\ell$ for $su(\ell)$, or for $\alpha$ arbitrary
for $so(\ell)$ and $sp(\ell)$ \cite{EHMM}.

An important consequence of the formula (\ref{def}) 
is that the spins of the
charges which it defines repeat modulo the Coxeter number $h$.
Consider first the case $\Lg = su(\ell)$, with $\alpha = 1/h$.
If $s/h$ is not an integer, then the expansion of $A(x,\lambda)^{s/h}$ is an
infinite series, and the expression for $\K_{s+1}$ will be
non-vanishing. 
If $s/h$ is an integer, however, then $A(x,\lambda)^{s/h}$ is 
a polynomial of degree $s$ in $\lambda$ and 
$\K_{s+1}$ vanishes according to the definition above.
Thus for $\Lg = su(\ell)$ the non-trivial charges have spins 
which run over all integers modulo the Coxeter number. 
For the algebras $\Lg = so(\ell)$ or $sp(\ell)$ things work in a more
trivial way, because $h$ is even, while the spins $s$ for the currents 
defined above are all the odd integers, so 
they certainly repeat modulo $h$.

So for each PCM based on a classical algebra we now have 
infinitely many commuting charges which come in sequences,
each sequence being associated with an exponent of the algebra and
with a corresponding primitive invariant tensor of type
(\ref{strace}).
But there is one primitive invariant tensor which is not of the type 
(\ref{strace}) and which has therefore been absent from our discussion
so far. This is the Pfaffian in $d_\ell = so(2 \ell)$.
It is natural to expect that our results can be extended to include
this last invariant, and this is indeed the case.

A direct computation with the first few examples listed in
(\ref{soN}) shows that $\int \K_m \, dx$ 
commutes with the Pfaffian charge $\int \p_\ell \, dx$ provided we 
choose $\alpha = 1/h$, where $h = 2\ell{-}2$ for $so(2\ell)$. 
But in fact the Pfaffian is just the first member 
of a series of conserved currents $\p_{\ell + ah}$ for 
integers $a \geq 0$ (where the subscript denotes the spin, as usual).
It is rather remarkable that these currents can be defined from the
same generating function $A(x,\lambda)$ we introduced above,
and using the same formula (\ref{def}), but with the coefficient to be
extracted as part of an expansion in {\em descending\/} powers of $\lambda$ 
rather than {\em ascending\/} powers of $\lambda$.
With this definition it can be shown that the charges 
$\int \K_m \, dx$ and $\int \p_{\ell + ah} \, dx$ all commute 
with one another for $\alpha = 1/h$. Full details 
are given in \cite{EHMM} .

In summary: in each PCM based on a classical algebra, 
we have found commuting local charges with spins equal to all the 
exponents modulo the Coxeter number.
We have not investigated PCMs based on 
exceptional groups, but we have no reason to suspect they should 
behave any differently.

\section{The classical supersymmetric PCM}

\subsection{The model in outline}

The most efficient way to supersymmetrize the bosonic PCM 
is to introduce a superfield
$G(x , \theta)$ with values in ${\g}$.
The additional coordinates are real Grassmann numbers $\theta^\pm$ 
with supercovariant derivatives 
$D_\pm = \partial_{\th^\pm} - i\th^\pm\partial_\pm$
which we can use to write a manifestly supersymmetric 
lagrangian\footnote{Spacetime conventions:  
Each index $\pm$ signifies one unit of Lorentz spin on a bosonic object, 
but a $1/2$-unit of spin on a fermionic object. 
Upper and lower indices denote opposite Lorentz weights.}
\be 
{\cal L}= {\rm Tr}( D_+ G^{-1} D_- G ) 
\ee
with ${\cal G}_L \times {\cal G}_R$ symmetry.
Corresponding to ${\cal G}_R$ and ${\cal G}_L$ we have the superspace currents 
\be 
J_\pm = - i G^{-1}D_\pm G 
\ee 
and $-G J_\pm G^{-1}$ respectively. 
These take values in $\Lg$ (tensored with the appropriate 
underlying real Grassmann algebra) and they satisfy 
\be\label{supercons}
D_+J_- = D_- J_+ = - {i\over 2} \{J_+, J_- \} \,.
\ee

Just as in the bosonic case, there are discrete symmetries 
$\pi: G \mapsto G^{-1}$ and $\gamma : G \mapsto G^*$ (and also
$\sigma : G \mapsto MGM^{-1}$ for $\Lg = so(2\ell)$).
We find 
$$
\pi \; : \; J_{++ \ldots +} \; \mapsto \; -G \,J_{++\ldots +} \,G^{-1} 
\; , \qquad
\gamma \; : \; 
J_{++ \ldots +} \; \mapsto \; (J_{++\ldots +})^*  = - (J_{++\ldots +})^T 
$$
where we have again introduced convenient quantities 
\be
J_{++}\equiv D_+ J_+\, +\, {i\over 2} \{J_+,J_+\} \; , \quad  
J_{+++}\equiv -i D_+ J_{++}\, +\, {1\over 2} [J_+,J_{++}] \; , \quad  
\ldots
\ee
which have simple behaviour.
Note that the brackets appearing above are graded, and that the factors 
of $i$ ensure that $J_{++\ldots+}$ is always a combination 
of anti-hermitian Lie algebra generators with real 
(possibly Grassmann algebra-valued) coefficients.

To reveal the component ($x$-space) content of the super PCM we 
can expand
$$ 
G(x , \theta) = g(x) ( 1 + i \theta^+ \psi_+ (x) + i
\theta^- \psi_- (x) + i \theta^+ \theta^- \sigma (x) \, ) \; .
$$
The fermions $\psi_\pm (x)$ take values in $\Lg$ and are the 
superpartners of the group-valued fields $g(x)$.
The field $\sigma(x)$ is auxiliary and can be eliminated from the 
action algebraically to produce four-fermion interaction terms.
The corresponding expansion of the superspace currents is 
\be\label{currcomp} 
J_\pm (x,\theta) = \psi_\pm(x) + \theta^\pm j_\pm(x) + \ldots 
\quad {\rm where} \quad j_\pm = - g^{-1} \del_\pm g - i \psi^2_\pm
\ee
and the precise form of the higher components of the currents 
will not be needed.
The equations of motion (\ref{supercons}) imply that the 
bosonic current is conserved, $ \del_- j_+ + \del_+ j_- = 0$,
although it does not obey (\ref{lce}).
The remaining consequences of (\ref{supercons}) 
are equations of motion for the fermions.

The classical super PCM is superconformally invariant, with the 
non-vanishing components of the super energy-momentum tensor
obeying 
$$
D_-{\rm Tr} (J_+ J_{++}) =
D_+{\rm Tr} (J_- J_{--}) =0 \ .
$$
When expanded in components this contains conservation equation for 
both the supersymmetry current and the conventional (bosonic) 
energy momentum tensor.

\subsection{Conserved charges}
The supersymmetric PCM contains infinitely many
local and non-local conserved quantities, some of which resemble their
bosonic counterparts, others arising in conjunction with
novel features. It has been known for a long time \cite{curt80} that 
the Yangian charges generalize to the supersymmetric theory 
with no significant modification of their properties. 
In particular, they commute with supersymmetry, so there is no
enhancement of Yangian symmetry.
We shall therefore concentrate on the local charges.

The simplest local conserved currents in the bosonic PCM are 
powers of the energy-momentum tensor (\ref{confpower}).
A super energy-momentum tensor is a fermionic quantity, however,
so we cannot take powers of it to obtain new conservation laws in
quite the same way. Let us therefore turn directly to the
generalizations of (\ref{curr}) and (\ref{gencons}).

\noindent
$\bullet$ The conservation laws (\ref{curr}) in the bosonic PCM
can be generalized to the supersymmetric PCM in two different ways.
First, we have 
\be\label{oddcurr}
D_- {\rm Tr}(J_+^{2n+1}) = 0 
\ee
which is odd under the discrete symmetry $\pi$.
The power of $J_+$ must be an odd integer, otherwise the expression 
would vanish identically, by Fermi statistics.
Second, we have 
\be \label{evencurr}
D_- {\rm Tr}(J_+^{2n-1} J^{\phantom{n}}_{++}) = 
0  
\ee
which is even under $\pi$. 
The power of $J_+$ must again be odd, this time to prevent the expression
being a total $D_+$ derivative and hence giving a trivial conservation
equation. Both (\ref{oddcurr}) and (\ref{evencurr}) 
follow directly from the superspace equations of motion.

\noindent
$\bullet$ As in the bosonic case, we can re-express and 
generalize these conservation equations by writing them in terms of
invariant tensors. The equation (\ref{oddcurr}) becomes 
\be\label{oddgen}
D_- ( \, \Omega_{a_1 a_2 \ldots a_{2n+1} } J^{a_1}_+ J^{a_2}_+ 
\ldots J^{a_{2n+1}}_+ \, ) = 0
\ee
where the odd-rank invariant tensor 
\be\label{omega} 
\Omega_{a_1 a_2 \ldots a_{2n+1}} =  
f_{ [ a_1 a_2}{}^{b_1} \ldots f_{a_{2n-1}
a_{2n}}{}^{b_n} d^{\,b_1 \ldots b_n}{}_{a_{2n+1} ]} 
\ee
is totally anti-symmetric.
In a similar fashion, the second kind of conservation equation 
(\ref{evencurr}) becomes 
\be\label{evengen}
D_- ( \, \Lambda_{a_1 \ldots a_{2n-1} a_{2n} } 
J_+^{a_1} \ldots J^{a_{2n-1}}_{+} J^{a_{2n}}_{++} \, ) =
0
\ee
where now the relevant invariant tensor is even-rank,
\be\label{lambda}
\Lambda_{a_1 a_2 \ldots a_{2n-1} a_{2n} }
= 
f_{ [ a_1 a_2}{}^{b_1} \ldots f_{a_{2n-3}
a_{2n-2}}{}^{b_{n-1}} d^{b_1 \ldots b_{n-1}}{}_{a_{2n-1} ] a_{2n}} 
\ee
It has a more complicated structure in that it is antisymmetric only
on its first $2n{-}1$ indices.

It seems natural that in a theory which contains fermionic
currents we should find conservation laws involving 
{\em antisymmetric\/} invariant tensors.
There can clearly only be finitely many of these.
We note that both $\Omega$ and $\Lambda$ are defined above 
in terms of some {\em symmetric\/} invariant tensor $d$.
They are non-vanishing when $d$ is one of the 
finite number of {\em primitive\/} symmetric tensors which we mentioned
previously (see {\it e.g.}~\cite{azca97}).

To get a better idea of the meaning of the above superspace 
conservation equations it is instructive to expand them in component
fields, using (\ref{currcomp}). 
On doing this we find that (\ref{oddgen}) produces fermionic and 
bosonic conserved currents 
\be\label{oddcomp} 
\Omega_{a_1 a_2 \ldots a_{2n+1} } \psi^{a_1}_+ \psi^{a_2}_+ 
\ldots \psi^{a_{2n+1}}_+
\; , \qquad 
\Omega_{a_1 \ldots a_{2n} a_{2n+1} } \psi^{a_1}_+ \ldots \psi^{a_{2n}}_+ 
j^{a_{2n+1}}_+
\ee
The fermionic and bosonic currents resulting from (\ref{evengen}) 
are more complicated. They can be written, up to terms proportional to
the expressions in (\ref{oddcomp}), as 
$$
d_{a_1 a_2 a_3 \ldots a_{n+1}} j_+^{a_1} \psi_+^{a_2} F_+^{a_3} \ldots 
F_+^{a_{n+1}} ,
\quad
d_{a_1 a_2 a_3 \ldots a_{n+1}} 
( nj_+^{a_1} j^{a_2}_+ + i \psi^{a_1}_+ \del_+
\psi^{a_2}_+ )  F_+^{a_3} \ldots F_+^{a_{n+1}}
$$
where we have introduced the bosonic quantity 
$F_+^a = if^a{}_{bc} \psi_+^b \psi_+^c$.
Notice that in either family of conservation laws, the fermionic and 
bosonic currents have spins $n{+}{1\over2}$ and $n{+}1$ respectively,
and so the corresponding conserved charges have spins
$n{-}{1\over2}$ and $n$ respectively. 
The $d$ tensors being primitive then implies  
that the values of $n$ are precisely the exponents of the algebra. 

The Poisson bracket structure of the super PCM and the resulting 
algebra of its local currents is significantly 
more complicated than in the bosonic case.
For this reason we shall not attempt to give a detailed discussion here. 
One can derive results for the families (\ref{oddcurr}) and 
(\ref{evencurr}) which are similar in many
respects to those we have described for the bosonic PCM.
The charges can be shown to have simple brackets 
amongst themselves, including many which vanish.
Finally, all of these charges commute with the Yangian. 
We intend to give a full account of these results 
in a forthcoming paper. 

\section{Remarks on quantum conserved charges}

The character of the bosonic and super PCMs changes dramatically 
on quantization. The (super)conformal invariance of the classical 
theories is broken, and the dimensionless classical coupling (which
we have suppressed throughout) is replaced by a mass-scale.
The theories are strongly coupled in the infra-red 
so that quantum computations from the classical action are 
usually formidable to say the least.

While the non-local charges have been successfully studied at the
quantum level, the situation for the local charges is more
complicated, and only indirect results are presently available.
To find some indication of whether the classical charges we have been
studying are also present in the quantum theory, we can use the 
method of Goldschmidt and Witten, summarized as follows.\footnote{The
Goldschmidt-Witten method in superspace was discussed in
\cite{clark81} and applied to the supersymmetric $O(N)$ sigma-model.}

\subsection{Goldschmidt-Witten counting} 

Suppose we have linearly-independent 
conservation equations $\del_- j_i = 0$ or $D_- J_i = 0$ (in the
supersymmetric case) with $i=1, \ldots , n$ and that these have a common 
prescribed behaviour under all symmetries of the theory. 
The only quantum modifications which can appear on the
right-hand sides of these equations are operators with the same mass 
dimension and the same behaviour under continuous and discrete symmetries.
Let $A_i$ with $i = 1 , \ldots , p$ be a linearly-independent set of
such operators.
We can also enumerate the linearly-independent 
total-derivative terms $B_i$ with $i = 1 , \ldots , q$ which 
have the same symmetry properties.
Since each of the $B$s is expressible 
as a combination of $A$s we must have $q\leq p$. Now 
if $n - p + q > 0 $, then there are at least this many 
combinations of the classical conservation equations which survive in
the quantum theory, because this is the number of linearly-independent
combinations for which the right-hand side is guaranteed to be a 
(super)spacetime divergence. 

We can now apply these arguments to the bosonic and 
supersymmetric PCMs, specializing to ${\cal G} = SU(\ell)$
for simplicity. It is important to consider the  
behaviour of each current under both the continuous symmetries and the 
discrete symmetries $\pi$ and $\gamma$. 
Starting with the bosonic model, it so happens that all the currents 
we list below have the {\em same} behaviour 
under $\pi$ and $\gamma$, and so we describe them simply as 
even or odd.

\noindent
$\bullet$ Spin-2: ${\rm Tr} (j_+^2)$, even; 
this is the energy-momentum tensor.
There is one anomaly $A_1= {\rm Tr}(j_- j_{++})$ and one derivative 
$B_1=\partial_+$Tr$(j_-j_+)$ with $A_1 = B_1$.
The conservation law therefore survives quantum-mechanically,
as we expect, but its modification reflects the non-vanishing of the 
trace of the quantum energy-momentum tensor, 
corresponding to the breaking of conformal symmetry.

\noindent
$\bullet$ Spin-3: ${\rm Tr}( j_+^3)$, odd.
There is one anomaly $A_1 = {\rm Tr}(j_{++}\{j_-,j_+\})$
and one derivative $B_1 = \partial_+{\rm Tr}(j_-j_+^2) $ with $A_1 =
B_1$; the conservation again survives quantization.

\noindent
$\bullet$ Spin-4: currents ${\rm Tr} (j^4_+)$ and $({\rm Tr}
(j^2_+))^2$ are both even under each of the discrete symmetries. 
The anomalies and derivatives with these symmetries are
\[
\begin{array}{lll}
A_1 = {\rm Tr}(j_-j_{++++}) & 
\hspace{1in}& B_1=\partial_+{\rm Tr}(j_-j_{+++}) \\[0.1in]
A_2 = {\rm Tr}(j_-j_+){\rm Tr}(j_+j_{++}) 
&& B_2=\partial_+ \left(
{\rm Tr}(j_-j_+){\rm Tr}(j_+^2) \right) \\[0.1in]
A_3 = {\rm Tr}(j_-j_{++}){\rm Tr}(j_+^2) 
&& B_3=\partial_+{\rm Tr}(j_-j_+^3)
\\[0.1in]
A_4 = {\rm Tr}(j_+^2\{j_-,j_{++}\}) 
&& B_4=\partial_-{\rm Tr}(j_{++}^2) \\[0.1in]
A_5 = {\rm Tr}(j_-j_+j_{++}j_+) && 
\end{array}
\]
Since $n=2$, $p=5$, $q=4$, we conclude that there is at least one 
linear combination of the currents which is conserved 
in the quantum theory.

\noindent
$\bullet$
For higher values of the spin, the Goldschmidt-Witten method is
inconclusive. 
For instance, we find for spin-5 (odd) that $n=2$, $p=8$, $q=6$; while for 
spin-6 (even) we have $n= 5$, $p=25$, $q=18$. 

Turning now to the supersymmetric $SU(\ell)$ PCM, we 
find the following results:-

\noindent
$\bullet$ Spin-3/2: ${\rm Tr}(J_+J_{++})$, even under both $\pi$ and 
$\gamma$; 
this is the super-energy-momentum tensor.
There is one anomaly $A_1={\rm Tr}(J_-J_{+++})$ and one derivative 
$B_1=D_+{\rm Tr}(J_-J_{++})$ with $A_1 = -B_1$.
Supersymmetry and translation invariance therefore survive in the
quantum theory, as expected.

\noindent
$\bullet$ Spin-3/2: ${\rm Tr}(J_+^3)$, odd under $\pi$, even under $\gamma$.
There is one anomaly $A_1={\rm Tr}(J_-[J_+,J_{++}])$ and one
derivative $B_1=D_+{\rm Tr}(J_-J_+^2)$ with $A_1=B_1$. This
current too survives. 

\noindent
$\bullet$ Spin-5/2: ${\rm Tr}(J_+^3J_{++})$, even under $\pi$, odd
under $\gamma$.
The lists of anomalies and derivatives with these symmetries are 
\[
\begin{array}{lll}
A_1 = {\rm Tr}(J_-\{J_+^2,J_{+++}\})& \hspace{1in}& 
B_1=D_+{\rm Tr}(J_+J_-J_+J_{++})\\[0.1in]
A_2 = {\rm Tr}(J_-J_+J_{+++}J_+)&& 
B_2=D_+{\rm Tr}(\{J_-,J_+^2\}J_{++})\\[0.1in]
A_3 = {\rm Tr}(\{J_-,J_+\}J_{++}^2)&& 
\\
\end{array}
\]
Since the former out-number the latter we do not 
necessarily have quantum conservation.

\noindent
$\bullet$ Spin-5/2: ${\rm Tr}(J_+^5)$, odd under both $\pi$ and $\gamma$.
This time we find 
\[
\begin{array}{lll}
A_1 ={\rm Tr}(J_-\{J_{++},J_{+++}\}) & \hspace{1in}& 
B_1=D_+{\rm Tr}(J_-J_{++}^2) \\[0.1in]

A_2= {\rm Tr}(J_-\{J_+,J_{++++}\}) && 
B_2=D_+{\rm Tr}(J_-[J_+,J_{+++}]) \\[0.1in]
A_3 ={\rm Tr}(\{J_-,J_+^3\}J_{++}) && 
B_3=D_+{\rm Tr}(J_-J_+^4) \\[0.1in]
A_4 ={\rm Tr}(J_+\{J_+,J_-\}J_+J_{++}) && 
B_4= D_-(J_+J_{++}^2) 
\end{array}
\]
and so this conservation law survives quantization. 

\noindent
$\bullet$
For higher values of the spin the results are inconclusive,
just as in the bosonic case. As illustrations, for spin-$7/2$ (odd/even) 
we have $n=2$, $p=28$, $q=20$; while 
for spin-$7/2$ (even/even) we find $n=2$, $p=27$, $q=20$.

\subsection{Implications of quantum conservation laws}

The counting arguments described above are sometimes sufficient to 
demonstrate the existence of a quantum conserved charge, but they are
by no means necessary.
The fact that they fail in most instances should certainly not be
interpreted as meaning that the classical equations in question do not
generalize, but merely that these arguments are insufficient to 
settle the matter one way or the other.
Moreover, it is believed that the existence of just one additional 
conserved charge of higher-spin---which the counting establishes 
for both the bosonic and super $SU(\ell)$ PCMs---is 
sufficient to guarantee 
integrability and factorization of the $S$-matrix. 
Since this in turn implies infinitely many more conserved quantities,
it would be somewhat surprising if, one charge 
being conserved, the others were not.

The survival of particular local charges which are even or odd under
some discrete symmetry (which we can call generically `parity')
can have important implications for the spectrum. 
Indeed, the contrasts between the local charges in the bosonic and
supersymmetric cases are reflected in the multiplets which
are required for the construction of consistent S-matrices 
\cite{ogie86,evans96}. 
In the bosonic case, odd-parity charges appear
only in conjunction with complex representations of $\Lg$,
and it is only such multiplets which form parity doublets.
In the supersymmetric case, the odd-parity family of currents
(\ref{oddgen}) is always present, 
matching nicely the assumptions of \cite{evans96},
where particle multiplets also appear in parity doublets.

Finally, we return to the main theme of our introductory remarks.
The quantum conservation of a full set of local charges 
with spins equal to the exponents would provide a natural 
explanation of the occurrence of Dorey's rule in the fusings of the PCM
S-matrices. This is immediate for the simply-laced cases, but there are
additional subtleties for the non-simply-laced theories, as
discussed in \cite{EHMM}. Their resolution is an interesting topic for
future work.

\noindent
{\bf Acknowledgments}:
We thank Jose Azc\'arraga, Patrick Dorey and G\'erard Watts for discussions.
JME is grateful to PPARC (UK) for an Advanced Fellowship.
NJM thanks Pembroke College Cambridge for a Stokes Fellowship, during
which much of this work was carried out. 
MH is grateful to St. John's College, Cambridge for a Studentship.
AJM thanks the Royal Commission of 1851 for a Research Fellowship.


{\small
\begin{thebibliography}{99}
\raggedright
%
\bibitem{EHMM}
J. M. Evans, H. Hassan, N. J. MacKay, A. J. Mountain, 
{\em Local conserved charges in principal chiral models\/};
{\tt hep-th/9902008}.
%
\bibitem{corri94}
E. Corrigan,
\newblock{\em Recent developments in affine Toda quantum field theory},
\newblock Lectures given at CRM-CAP Summer School on Particles and
Fields '94, Banff, Canada, 16-24 Aug 1994,
\newblock preprint DTP-94/55; {\tt hep-th/9412213}.
%
\bibitem{brade90}
H. W. Braden, E. Corrigan, P. E. Dorey and R. Sasaki,
{\em Affine Toda field theory and exact S-matrices}, Nucl. Phys. {\bf
B338} (1990) 689.
%
\bibitem{dorey91} P. E. Dorey, {\em Root systems and purely elastic
S-matrices}, Nucl. Phys. {\bf B358} (1991) 654.
%
\bibitem{ganden96}
G. M. Gandenberger, {\em Exact $S$-matrices for quantum affine Toda
solitons and their bound states}, Ph.D.\ thesis, Cambridge University
1996, available at 
{\tt http://www.damtp.cam.ac.uk/user/hep/publications.html}.
%
\bibitem{ogie86}
E. Ogievetsky, N. Reshetikhin and P. Wiegmann, {\em The principal
chiral field in two dimensions on classical Lie algebras: the Bethe
ansatz solution and factorized theory of scattering}, 
Nucl. Phys. {\bf B280} (1987) 45.

E. Abdalla, M. C. B. Abdalla and A. Lima-Santos,
{\em On the exact S-matrix of the principal chiral model}, 
Phys.~Lett.~{\bf B140} (1984) 71, erratum {\bf B146} (1984) 457.
%
\bibitem{evans96}
J. M. Evans and T. J. Hollowood, {\em Exact scattering in the $SU(N)$
supersymmetric principal chiral model},
Nucl. Phys. {\bf B493} (1997) 517; {\tt hep-th/9603190}.
%
\bibitem{TBA}
J. M. Evans and T. J. Hollowood, {\em Exact results for integrable
asymptotically-free field theories\/} Nucl. Phys. {\bf B}
(Proc.~Suppl.) {\bf 45A} (1996) 130; {\tt hep-th/9508141}.
%
\bibitem{qyang}
M. L\"uscher, {\em Quantum non-local charges and the absence of particle
production in the 2D non-linear $\sigma$-model}, Nucl. Phys. {\bf B135}
(1978) 1.

D. Bernard,
{\em Hidden Yangians in 2D massive current algebras}, Commun.~Math.~Phys. 
{\bf 137} (1991) 191.
%
\bibitem{cyang}
E. Br\'ezin, C. Itzykson, J. Zinn-Justin and J.-B. Zuber,
{\em Remarks on the existence of non-local charges in two-dimensional
models}, Phys. Lett. {\bf 82B} (1979) 442.

M. L\"uscher and K. Pohlmeyer, {\em Scattering of massless lumps
and non-local charges in the two-dimensional classical non-linear
sigma model}, Nucl. Phys. {\bf B137} (1978) 46.

N. J. MacKay, {\em On the classical origins of Yangian symmetry 
in integrable field theory}, Phys. Lett. {\bf B281} (1992) 90;
err. ibid. {\bf B308} (1993) 444.
%
\bibitem{chari95} 
V. Chari and A. Pressley, {\em Yangians, integrable quantum systems
and Dorey's rule}, Commun. Math. Phys. {\bf 181} (1996) 265;
{\tt hep-th/9505085}.
%
\bibitem{azca97}
J. A. de Azcarraga, A. J. MacFarlane, A. J. Mountain and 
J. C. P\'erez Bueno, {\em Invariant tensors for simple groups},
Nucl. Phys. {\bf B510} (1998) 657; {\tt physics/9706006}.

A. J. Mountain, 
{\em Invariant tensors and Casimir operators for simple compact Lie groups\/},
J.~Math Phys.~{\bf 39} (1998) 5601; {\tt physics/9802012}.
%
\bibitem{fadd86}
L. Faddeev and N. Reshetikhin,
{\em Integrability of the principal chiral field in 1+1 dimensions},
Ann. Phys. (NY) {\bf 167} (1986) 227.
%
\bibitem{curt80}
T. Curtright and C. Zachos, {\em Nonlocal currents for supersymmetric
nonlinear models}, Phys. Rev. {\bf D21} (1980) 411.

E. Corrigan and C. Zachos, {\em Non-local charges for the
supersymmetric $\sigma$-model}, Phys. Lett. {\bf B88} (1979) 273.
%
\bibitem{gold80}
Y. Y. Goldschmidt and E. Witten, {\em Conservation laws in
some two-dimensional models}, Phys. Lett. {\bf B91} (1980) 392.

E. Witten, {\em Lectures on Field Theory}, no.\ 3, 
Princeton QFT program (1996-7), available at 
{\tt http://www.math.ias.edu/$\sim$drm/QFT/}.
%
\bibitem{clark81}
T. E. Clark, S. T. Love and S. Gottlieb,
{\em Infinite number of conservation laws in two-dimensional
superconformal models},
Nucl. Phys. {\bf B186} (1981) 347.
\end{thebibliography}
}
\end{document}